\documentclass[12pt]{article}

\usepackage{amsfonts}
\usepackage{amssymb}
\usepackage{amscd}

\def\D{\hbox{D\kern-.73em\raise.25ex\hbox{-}\raise-.25ex\hbox{ }}}
 \def\d{\hbox{d\kern-.33em\raise.75ex\hbox{-}\raise-.75ex\hbox{}}}

\def\GGG{\frak G }
\def\gr3{\GGG\,(\SSS_3)}

\def\gr2{\GGG\,(\SSS_2)}

\def\SSS{\frak S}

\def\ed{\end{document}}
\def\beq{\begin{equation}}
\def\eeq{\end{equation}}
\def\bea{\begin{eqnarray}}
\def\eea{\end{eqnarray}}
\def\ba{\begin{array}}
\def\ea{\end{array}}
\def\bi{\begin{itemize}}
\def\ei{\end{itemize}}

\newcommand{\bp}{\noindent\begin{minipage}[c]}
\newcommand{\ep}{\end{minipage}}

\begin{document}
\title{\bf  Some Lagrangians with \\ Zeta Function Nonlocality  }

\author{Branko Dragovich\thanks{\,e-mail
address: dragovich@phy.bg.ac.yu} \\ {}\\
\it{Institute of Physics}\\ \it{Pregrevica 118, P.O. Box 57, 11001
Belgrade, Serbia}}

\date {~}
\maketitle
\begin{abstract}
Some  nonlocal and nonpolynomial scalar field models originated from
$p$-adic string theory are considered. Infinite number of spacetime
derivatives is governed by the   Riemann zeta function through
d'Alembertian $\Box$ in its argument. Construction of the
corresponding Lagrangians  begins with the exact Lagrangian  for
effective field of $p$-adic tachyon string, which is generalized
replacing $p$ by arbitrary natural number $n$ and then taken a sum
of  over all $n$. Some basic classical field properties of these
scalar fields are obtained. In particular, some trivial solutions of
the equations of motion and their tachyon spectra are presented.
Field theory with Riemann zeta function nonlocality is also
interesting in its own right.
\end{abstract}
\bigskip

{\it \hskip5cm Dedicated to Professor I.L. Buchbinder}

{\it \hskip5cm on the occasion of his 60th anniversary}

\section{Introduction}

 The first paper on a $p$-adic string is published in 1987
\cite{volovich1}.
 After that various $p$-adic structures have been observed not only in
string theory but also in many other models of modern mathematical
physics (for a review of the early days developments, see e.g.
\cite{freund1,volovich2}).

One of the remarkable achievements in $p$-adic string theory is
construction of a field model for open scalar $p$-adic string
\cite{freund2,frampton1}. The effective tachyon Lagrangian is very
simple and exact. It describes  four-point scattering amplitudes as
well as  all higher ones at the tree-level.

This field theory approach to $p$-adic string theory  has been
significantly pushed forward when was shown \cite{sen} that it may
describes tachyon condensation and brane descent relations. After
this success, many aspects of $p$-adic string dynamics have been
investigated and compared with dynamics of ordinary strings (see,
e.g. \cite{minahan,sen1,zwiebach,arefeva1} and references therein).
Noncommutative deformation of $p$-adic string world-sheet with a
constant B-field was  investigated in \cite{ghoshal-grange} (on
$p$-adic noncommutativity see also \cite{dragovich1}). A systematic
mathematical study of spatially homogeneous solutions of the
relevant nonlinear differential equations of motion has been of
considerable interest (see \cite{zwiebach,vladimirov1,vladimirov2,
barnaby1} and references therein). Some possible cosmological
implications of $p$-adic string theory have been also investigated
\cite{arefeva2,arefeva3,barnaby2,arefeva4,calcagni}. It was proposed
\cite{ghoshal} that $p$-adic string theories provide lattice
discretization to the world-sheet of ordinary strings. As a result
of these developments, some nontrivial features of ordinary string
theory have been reproduced from the $p$-adic effective action.
Moreover, there have been established  many similarities and
analogies between $p$-adic and ordinary strings.

Adelic approach to the string scattering amplitudes enables to
connect $p$-adic and ordinary counterparts (\cite{freund1,
volovich2} as a review, and see also \cite{branko1}). Moreover, it
eliminates unwanted prime number parameter $p$ contained in $p$-adic
amplitudes and also cures the problem of $p$-adic causality
violation. Adelic generalization of quantum mechanics was also
successfully formulated, and it was found a connection between
adelic vacuum state of the harmonic oscillator and the Riemann zeta
function \cite{dragovich2}. Recently, an interesting approach toward
foundation of a field theory and cosmology based on the Riemann zeta
function was proposed in \cite{volovich3}. Note that $p$-adic and
ordinary sectors of the four point adelic string amplitudes
separately contain the Riemann zeta function (see, e.g.
\cite{freund1}, \cite{volovich2} and \cite{dragovich4}).

The present paper is mainly motivated by our intention to obtain the
corresponding effective Lagrangian for  adelic scalar string. Hence,
as a first step we investigate possibilities to derive Lagrangian
related to the $p$-adic sector of adelic string. Starting with the
exact Lagrangian for the effective field of $p$-adic tachyon string,
extending prime number $p$  to arbitrary natural number $n$ and
undertaking various summations of such Lagrangians over all $n$, we
obtain some scalar field models with the operator valued Riemann
zeta function. Emergence of the Riemann zeta function at the
classical level can be regarded as its analog  of quantum scattering
amplitude. This zeta function controls spacetime nonlocality. In the
sequel  we shall construct and explore some classical field models
which should help in investigation of some properties of  adelic
scalar strings.

\section{Construction of  zeta nonlocal Lagrangians}

The exact tree-level Lagrangian of effective scalar field $\varphi$
for open $p$-adic string tachyon is

\beq {\cal L}_p = \frac{m_p^D}{g_p^2}\, \frac{p^2}{p-1} \Big[
-\frac{1}{2}\, \varphi \, p^{-\frac{\Box}{2 m_p^2}} \, \varphi  +
\frac{1}{p+1}\, \varphi^{p+1} \Big]\,,  \label{2.1} \eeq where $p$
 is any prime number, $\Box = - \partial_t^2  + \nabla^2$ is the
$D$-dimensional d'Alembertian.

The equation of motion for (\ref{2.1}) is

\beq p^{-\frac{\Box}{2 m_p^2}}\, \varphi = \varphi^p \,, \label{2.2}
\eeq and its properties have been studied by many authors (see e.g.
\cite{zwiebach,vladimirov1,vladimirov2,barnaby1} and references
therein).

Prime number $p$ in (\ref{2.1}) and (\ref{2.2}) can be replaced by
any natural number $n \geq 2$ and such expressions also make sense.
Moreover,  if $p = 1 + \varepsilon \to 1$  there is the limit of
(\ref{2.1})

\beq {\cal L} = \frac{m^D}{g^2}\,  \Big[ \frac{1}{2}\, \varphi \,
\frac{\Box}{m^2} \, \varphi  + \frac{\varphi^2}{2}\, ( \ln
\varphi^{2} -1 ) \Big]\, \label{2.3}\eeq
 which corresponds to the ordinary bosonic string in the boundary string field
theory \cite{gerasimov}.

Now we want to introduce a model which incorporates all the above
string Lagrangians (\ref{2.1}) with $p$ replaced by $n \in
\mathbb{N}$. To this end, we take the sum of all Lagrangians ${\cal
L}_n$  in the form

\bea L =   \sum_{n = 1}^{+\infty} C_n\, {\cal L}_n   =  \sum_{n=
1}^{+\infty} C_n \frac{ m_n^D}{g_n^2}\frac{n^2}{n -1} \Big[
-\frac{1}{2}\, \phi \, n^{-\frac{\Box}{2 m_n^2}} \, \phi +
\frac{1}{n + 1} \, \phi^{n+1} \Big]\,, \label{2.4} \eea whose
explicit realization depends on particular choice of coefficients
$C_n$, string masses $m_n$ and coupling constants $g_n$. To avoid a
divergence problem in  $1/(n-1)$ when $n = 1$ one has to take that
${C_n\, m_n^D}/{g_n^2}$ is proportional to $n -1$. In this paper we
shall consider a case when coefficients $C_n$ are proportional to
$n-1$, while masses $m_n$ as well as coupling constants $g_n$ do not
depend on $n$, i.e. $ m_n =  m , \,\, g_n = g$. Since this is an
approach towards effective Lagrangian  of an adelic string it seems
natural  to take mass and coupling constant independent on
particular $p$ or $n$. To emphasize that Lagrangian (\ref{2.4})
describes a new field, which is different from a particular $p$-adic
one, we introduced notation $\phi$ instead of $\varphi$. The two
terms in (\ref{2.4}) with $n = 1$ are equal up to the sign, but we
remain them because they provide the suitable form of total
Lagrangian $L$.

\subsection{Case $C_n = \frac{n-1}{n^{2+h}}$ }

Let us first consider  the case

\beq C_n = \frac{n-1}{n^{2+h}} \,, \label{2.5} \eeq where $h$ is a
real number. The corresponding Lagrangian is

\bea L_{h} =   \frac{m^D}{g^2} \Big[ - \frac{1}{2}\, \phi \,
\sum_{n= 1}^{+\infty} n^{-\frac{\Box}{2 m^2} -h} \, \phi  + \sum_{n=
1}^{+\infty} \frac{n^{-h}}{n + 1} \, \phi^{n+1} \Big] \label{2.6}
\eea and it depends on parameter $h$.

According to the famous Euler product formula one can write
$$ \sum_{n= 1}^{+\infty} n^{-\frac{\Box}{2\, m^2}  - h} = \prod_p \frac{1}{ 1 -
p^{-\frac{\Box}{2\, m^2}   - h}}\,. $$  Recall that standard
definition of the Riemann zeta function is

\beq  \zeta (s) = \sum_{n= 1}^{+\infty} \frac{1}{n^{s}} = \prod_p
\frac{1}{ 1 - p^{- s}}\,, \quad s = \sigma + i \tau \,, \quad \sigma
>1\,, \label{2.7} \eeq which has analytic continuation to the entire
complex $s$ plane, excluding the point $s=1$, where it has a simple
pole with residue 1. Employing definition (\ref{2.7}) we can rewrite
(\ref{2.6}) in the form

 \beq L_{h} = \frac{m^D}{g^2} \Big[ \, -
\frac{1}{2}\,
 \phi \,  \zeta\Big({\frac{\Box}{2\, m^2}  +
h }\Big) \, \phi  +   \sum_{n= 1}^{+\infty} \frac{n^{ - h}}{n + 1}
\, \phi^{n+1} \Big]\,. \label{2.8} \eeq Here
 $\zeta\Big({\frac{\Box}{2\, m^2} + h}\Big)$ acts as a
pseudodifferential operator  \beq \label{2.9}
\zeta\Big({\frac{\Box}{2\, m^2}  + h }\Big)\, \phi (x) =
\frac{1}{(2\pi)^D}\, \int e^{ ixk}\, \zeta\Big(-\frac{k^2}{2\, m^2}
 + h \Big)\, \tilde{\phi}(k)\,dk \,,
 \eeq
where $ \tilde{\phi}(k) =\int e^{(- i kx)} \,\phi (x)\, dx$ is the
Fourier transform of $\phi (x)$. Lagrangian $L_0 $, with the
restriction on momenta $-k^2 = k_0^2 -\overrightarrow{k}^2
> (2 - 2 h)\, m^2 $ and field $|\phi | < 1$, is analyzed in \cite{dragovich3}. In the
sequel we shall consider Lagrangian (\ref{2.8}) with analytic
continuations of the zeta function and the power series $\sum
\frac{n^{-h}}{n + 1} \, \phi^{n+1}$, i.e.

\beq L_{h} = \frac{m^D}{g^2} \Big[ \,- \frac{1}{2}\,
 \phi \,  \zeta\Big({\frac{\Box}{2 \, m^2}  +
h }\Big) \, \phi    + {\cal{AC}} \sum_{n= 1}^{+\infty} \frac{n^{-
h}}{n + 1} \, \phi^{n+1} \Big]\,, \label{2.10} \eeq where
$\mathcal{AC}$ denotes analytic continuation.

 Nonlocal dynamics of this field $\phi$ is
encoded in the pseudodifferential form of the Riemann zeta function.
When the d'Alembertian is in the argument of the Riemann zeta
function we  say that we have zeta nonlocality. Accordingly, this
$\phi$ is a zeta nonlocal scalar field.

Potential of the above zeta scalar field (\ref{2.10}) is equal to $-
L_h$ at $\Box = 0$, i.e.

\beq V_{h} (\phi) = \frac{m^D}{ g^2}\,\Big( \frac{\phi^2}{2} \,
\zeta ( h)   - \mathcal{AC} \sum_{n= 1}^{+\infty} \frac{n^{ - h}}{n
+ 1}\, \phi^{n +1} \Big)\,, \label{2.11}\eeq where $h \neq 1$ since
$\zeta (1) = \infty$. The term with $\zeta$-function vanishes at $h
= -2, -4, -6, \cdots$.

The equation of motion in differential and integral form is

\bea  \zeta\Big({\frac{\Box}{2\, m^2}  + h }\Big) \, \phi
= \mathcal{AC} \sum_{n = 1}^{+\infty} n^{- h}\, \phi^n \,, \\
\frac{1}{(2\pi)^D}\, \int_{\mathbb{R}^D} e^{ ixk}\,
\zeta\Big(-\frac{k^2}{2 \, m^2}  + h \Big)\, \tilde{\phi}(k)\,dk =
\mathcal{AC} \sum_{n = 1}^{+\infty} n^{ - h}\, \phi^n \,,
\label{2.12} \eea respectively. It is clear that $\phi =0$ is a
trivial solution for any real $h$. Existence of other trivial
solutions depends on parameter $h$. When $h > 1$ we have another
constant trivial solution $\phi = 1$.

In the weak field approximation $(|\phi (x)|\ll 1)$ the above
expression (\ref{2.12}) becomes

\beq  \int_{\mathbb{R}^D} e^{i k x} \, \Big[\zeta\Big(-\frac{ k^2}{2
\, m^2} + h \Big) - 1 \Big]\, \tilde{\phi}(k)\, dk = 0 \,,
\label{2.13} \eeq which has a solution $\tilde{\phi}(k) \neq 0$ if
equation

\beq \zeta\Big( \frac{-k^2}{2\, m^2} + h \Big) = 1 \label{2.14} \eeq
is satisfied. According to the usual relativistic kinematic relation
$k^2 = - k_0^2 +\overrightarrow{k}^2 = - M^2$, equation (\ref{2.14})
in the form

\beq \zeta\Big( \frac{M^2}{2\, m^2} + h \Big) = 1\,, \label{2.15}
\eeq determines mass spectrum $ M^2 = \mu_h \, m^2$, where set of
values of spectral function $\mu_h$ depends on $h$.

Equation (\ref{2.15}) gives infinitely many tachyon  mass solutions.
Namely, function $\zeta (s)$ is continuous for real $s \neq 1$ and
changes sign crossing its zeros  $s = - 2n, \, n \in \mathbb{N}$.
 According to relation $\zeta (1 - 2n) = - B_{2n}/ (2n)$ and values
 of the Bernoulli numbers $(B_0 = 1, \, B_1 = - 1/2, \, B_2 =  1/6, \, B_4 = - 1/30, \,
 B_6 = 1/42, \, B_8 = - 1/30, \, B_{10} =  5/66, \, B_{12} = - 691/2730, \, B_{14} =  7/6, \,
 B_{16} = - 3617/510, \, B_{18} = 43867/798, \, \cdots )$ it follows
 that $|\zeta (1 - 2n)| = | B_{2n}/ (2n)| > 1$ if and only
 if $n \geq 9$. Taking into account also regions where $\zeta (1 -2n)  >
 0$ we conclude that $\zeta (s) =1$ has two solutions  when $  -20 - 4j < s < - 18 - 4j
 $ for every $j = 0, 1, 2, \cdots$. Consequently, for any $h \in \mathbb{Z}$, we obtain
 infinitely many tachyon masses $ M^2$:
 \beq M^2 = - (40 + 8 j + 2 h - a_j)\, m^2 \quad \mbox{and} \quad M^2 = - (36 + 8 j + 2 h + b_j)\, m^2 , \label{2.16}  \eeq
where $a_j \ll 1$, $b_j \ll 1$ and $j = 0, 1, 2, \cdots$.

An elaboration  of the above Lagrangian for $h = 0, \pm 1, \pm 2$ is
presented in \cite{dragovich5}.

\subsection{Case $C_n = \frac{n^2 - 1}{n^2}$ }

In this case Lagrangian (\ref{2.4}) becomes

\bea L =   \frac{m^D}{g^2} \Big[ - \frac{1}{2}\, \phi \, \sum_{n=
1}^{+\infty} \Big( n^{-\frac{\Box}{2 m^2} + 1} \, + \,
n^{-\frac{\Box}{2 m^2}} \Big) \, \phi  + \sum_{n= 1}^{+\infty} \,
\phi^{n+1} \Big] \label{2.17} \eea and it yields

 \beq L = \frac{m^D}{g^2} \Big[ \, - \frac{1}{2}\,
 \phi \,  \Big\{ \zeta\Big({\frac{\Box}{2\, m^2}  -
 1}\Big)\, + \, \zeta\Big({\frac{\Box}{2\, m^2} }\Big) \Big\} \, \phi \,  + \,   \frac{\phi^2}{1 - \phi} \,
 \Big]\,. \label{2.18} \eeq

The corresponding potential is

 \beq  V (\phi) = - \frac{m^D}{g^2}  \,   \frac{31 - 7 \phi}{24\, (1 - \phi)}
 \,\phi^2
 \,, \label{2.19} \eeq
 which has the following properties: $V (0) = V (31/7) = 0\,, \,\, V (1\pm 0) = \pm \infty\,, \,\, V (\pm \infty) = -
 \infty$. At $\phi = 0$ potential has local maximum.

The equation of motion is

\beq \Big[ \zeta\Big({\frac{\Box}{2\, m^2}  -
 1}\Big)\, + \, \zeta\Big({\frac{\Box}{2\, m^2} }\Big) \Big] \, \phi
 = \frac{\phi ((\phi - 1)^2 + 1)}{(\phi - 1)^2}\,, \label{2.20}
\eeq which has only  $\phi = 0$ as a constant real solution. Its
weak field approximation is

\beq \Big[ \zeta\Big({\frac{\Box}{2\, m^2}  -
 1}\Big)\, + \, \zeta\Big({\frac{\Box}{2\, m^2} }\Big) - 2 \Big] \, \phi
 = 0\,, \label{2.21}
\eeq which implies condition on the mass spectrum

\beq  \zeta\Big(\frac{M^2}{2\, m^2}  -
 1\Big)\, + \, \zeta\Big({\frac{M^2}{2\, m^2} }\Big)
 = 2\,. \label{2.22}\eeq
From (\ref{2.22}) it follows one solution for $M^2 > 0$ at $M^2
\thickapprox 2.79\, m^2$ and many tachyon solutions when $M^2 < - 38
\, m^2 $.

\section{Extension by ordinary Lagrangian}

Let us now add ordinary bosonic Lagrangian (\ref{2.3}) to the above
constructed ones, i.e. $\mathbf{L}_h = L_h + \mathcal{L}$ and
$\mathbf{L} = L + \mathcal{L}$.

Respectively, one has Lagrangian, potential, equation of motion and
mass spectrum condition:

\beq \mathbf{L}_h = \frac{m^D}{g^2} \Big[ \frac{\phi}{2}\, \Big\{
 \frac{\Box}{ m^2} - \zeta\Big(\frac{\Box}{2\, m^2} + h \Big)\Big\}\phi + \frac{\phi^2}{2}\, (\ln{\phi^2} -1  )
 + \mathcal{A C} \sum_{n=1}^{+ \infty} \frac{n^{- h}}{n +1} \phi^{n+1} \Big] \,, \label{2.23} \eeq

\beq \mathbf{V}_h (\phi) = \frac{m^D}{g^2} \Big[ \frac{\phi^2}{2}\,
\Big(\zeta (h) + 1 - \ln{\phi^2}\Big) - \mathcal{A C} \sum_{n=1}^{+
\infty} \frac{n^{-h}}{n +1}\, \phi^{n+1} \Big] \,, \label{2.24}
\eeq

\beq \Big[ \zeta\Big(\frac{\Box}{2\, m^2} + h \Big) - \frac{\Box}{
m^2} \Big]\, \phi = \phi\, \ln{\phi^2} + \mathcal{A C} \sum_{n=1}^{+
\infty} \frac{\phi^n}{n^h} \,, \label{2.25} \eeq

\beq \zeta\Big(\frac{M^2}{2\, m^2} + h \Big) - \frac{M^2}{ m^2} = -
1 \,. \label{2.26} \eeq

An analysis of these expressions depending on parameter $h$ will be
presented elsewhere. When $C_n = \frac{n^2 - 1}{n^2}$ one
respectively obtains:

\beq \mathbf{L} = \frac{m^D}{g^2} \Big[ \frac{\phi}{2}\, \Big\{
 \frac{\Box}{ m^2} - \zeta\Big(\frac{\Box}{2\, m^2} - 1 \Big) - \zeta\Big(\frac{\Box}{2\, m^2}  \Big)
 - 1 \Big\}\phi + \frac{\phi^2}{2}\, \ln{\phi^2} +
  \frac{\phi^2}{1- \phi} \Big] \,, \label{2.27} \eeq

\beq \mathbf{V} (\phi) = \frac{m^D}{g^2}  \frac{\phi^2}{2}\,
\Big[\zeta (-1) + \zeta (0) + 1 - \ln{\phi^2} - \frac{1}{1- \phi}
\Big] \,, \label{2.28} \eeq

\beq \Big[ \zeta\Big(\frac{\Box}{2\, m^2} -1 \Big) + \zeta \Big(
\frac{\Box}{2\, m^2} \Big) -  \frac{\Box}{ m^2} + 1 \Big]\, \phi =
\phi\, \ln{\phi^2} + \phi + \frac{2 \phi - \phi^2}{(1 - \phi)^2} \,,
\label{2.29} \eeq

\beq \zeta\Big(\frac{M^2}{2\, m^2} -1 \Big) +
\zeta\Big(\frac{M^2}{2\, m^2} \Big) = \frac{M^2}{m^2} \,.
\label{2.30} \eeq

Potential (\ref{2.28}) has one local minimum $\mathbf{V}(0) =0$ and
two local maxima, which are approximately: $\mathbf{V}(-0.6)
\thickapprox 0.15 \,\frac{m^D}{g^2}$ and $\mathbf{V}(0.3)
\thickapprox 0.06 \,\frac{m^D}{g^2}$. It has also the following
properties: $\mathbf{V}(1\pm 0) = \pm \infty$ and $\mathbf{V}(\pm
\infty) = - \infty$.

In addition to many tachyon solutions, equation (\ref{2.30}) has two
solutions with positive mass: $M^2 \thickapprox 2.67\, m^2$ and $M^2
\thickapprox 4.66\, m^2$.

\section{Concluding remarks}
As a first step towards construction of an effective field theory
for  adelic open scalar string, we have found a few  Lagrangians
which contain all corresponding $n$-adic Lagrangians ($n \in
\mathbb{N}$). As a result one obtains that an infinite number of
spacetime derivatives and related nonlocality are governed by the
Riemann zeta function. Potentials are nonpolynomial. Tachyon mass
spectra are determined by definite equations and they are contained
in all the above cases. $p$-Adic Lagrangians can be easily restored
from a zeta Lagrangian using just an inverse procedure for its
construction.

This paper contains  some basic classical  properties of the
introduced scalar field with zeta function nonlocality. There are
rather many classical aspects which should  be investigated. One of
them is a systematic study of the equations of motion and nontrivial
solutions. In the quantum sector it is  desirable to investigate
scattering amplitudes and make comparison with adelic string.

\section*{\large Acknowledgements}
The work on this article was partially supported by the Ministry of
Science, Serbia, under contract No 144032D. The author thanks I. Ya.
Aref'eva and I. V. Volovich for useful discussions. This paper was
completed during author's stay in the Steklov Mathematical
Institute, Moscow.

 \end{document}